\documentclass[iop]{emulateapj}

\def\D{\Delta}

\newcommand{\target}{K2-34}
\usepackage{amsmath,amssymb}
\usepackage{graphicx}
\usepackage{natbib}

\usepackage{aas_macros} 
\slugcomment{}
\shortauthors{Hirano et al.}
\shorttitle{The K2-ESPRINT Project IV}
\begin{document}
\title{The K2-ESPRINT Project IV: A Hot Jupiter in a Prograde Orbit with a
Possible Stellar Companion}
\author{
Teruyuki Hirano\altaffilmark{1}, 
Grzegorz Nowak\altaffilmark{2,3},
Masayuki Kuzuhara\altaffilmark{1,6},
Enric Palle\altaffilmark{2,3},
Fei Dai\altaffilmark{4},
Liang Yu\altaffilmark{4},
Vincent Van Eylen\altaffilmark{5},
Yoichi Takeda\altaffilmark{6}, 
Timothy D. Brandt\altaffilmark{7,8},
Norio Narita\altaffilmark{6,9,10}, 
Sergio Velasco\altaffilmark{2,3},
Jorge Prieto Arranz\altaffilmark{2,3},
Roberto Sanchis-Ojeda\altaffilmark{8,11},
Joshua N. Winn\altaffilmark{4},
Tomoyuki Kudo\altaffilmark{12},
Nobuhiko Kusakabe\altaffilmark{10},
Akihiko Fukui\altaffilmark{13}, 
Bun'ei Sato\altaffilmark{1}, 
Simon Albrecht\altaffilmark{5},
Ignasi Ribas\altaffilmark{14},
Tsuguru Ryu\altaffilmark{9},
Motohide Tamura\altaffilmark{6,8,15}
} 

\altaffiltext{1}{Department of Earth and Planetary Sciences, Tokyo Institute of Technology, 2-12-1 Ookayama, Meguro-ku, Tokyo 152-8551, Japan}
\email{hirano@geo.titech.ac.jp}
\altaffiltext{2}{Instituto de Astrof\'{i}sica de Canarias (IAC), 38205 La Laguna, Tenerife, Spain}
\altaffiltext{3}{Departamento de Astrof\'{i}sica, Universidad de La Laguna (ULL), 38206 La Laguna, Tenerife, Spain}
\altaffiltext{4}{Department of Physics, and Kavli Institute for Astrophysics and Space Research, Massachusetts Institute of Technology, Cambridge, MA 02139}
\altaffiltext{5}{Stellar Astrophysics Centre, Department of Physics and Astronomy, Aarhus University, Ny Munkegade 120, DK-8000 Aarhus C, Denmark}
\altaffiltext{6}{National Astronomical Observatory of Japan, 2-21-1 Osawa, Mitaka, Tokyo 181-8588, Japan}
\altaffiltext{7}{Astrophysics Department, Institute for Advanced Study, Princeton, NJ, USA}
\altaffiltext{8}{NASA Sagan Fellow}
\altaffiltext{9}{SOKENDAI (The Graduate University for Advanced Studies), 2-21-1 Osawa, Mitaka, Tokyo 181-8588, Japan}
\altaffiltext{10}{Astrobiology Center, National Institutes of Natural Sciences, 2-21-1 Osawa, Mitaka, Tokyo 181-8588, Japan}
\altaffiltext{11}{Department of Astronomy, University of California, Berkeley, CA 94720}
\altaffiltext{12}{Subaru Telescope, National Astronomical Observatory of Japan, 650 North Aohoku Place, Hilo, HI 96720, USA}
\altaffiltext{13}{Okayama Astrophysical Observatory, National Astronomical Observatory of Japan, Asakuchi, Okayama 719-0232, Japan}
\altaffiltext{14}{Institut de Ci\`{e}ncies de l'Espai (CSIC-IEEC), Carrer de Can Magrans, Campus UAB, 08193 Bellaterra, Spain}
\altaffiltext{15}{Department of Astronomy, Graduate School of Science, The University of Tokyo, Hongo 7-3-1, Bunkyo-ku, Tokyo, 113-0033}

\begin{abstract}
We report on the detection and early characterization of a hot Jupiter 
in a 3-day orbit around \target\ (EPIC~212110888), a metal-rich F-type star located 
in the K2 Cycle 5 field. 
Our follow-up campaign involves precise radial velocity (RV) measurements
and high-contrast imaging using multiple facilities. 
The absence of a bright nearby source in our high-contrast data suggests that the transit-like signals are not due to light variations from such a companion star.
Our intensive RV measurements show that \target b has a mass of $1.773\pm0.086M_J$, confirming its status as a planet.
We also detect the Rossiter-McLaughlin effect for \target b and show
that the system has a good spin-orbit alignment ($\lambda=-1_{-9}^{+10}$ degrees). 
High-contrast images obtained by the HiCIAO camera on the Subaru 8.2-m
telescope reveal a faint companion candidate ($\Delta m_H=6.19\pm 0.11$ mag) at a separation of $0\farcs36$. Follow-up observations are needed to confirm that the companion candidate is physically associated with \target. 
\target b appears to be an example of a typical ``hot Jupiter,'' albeit one which can be precisely characterized using a combination of K2 photometry and ground-based follow-up.
\end{abstract}

\section{Introduction}\label{s:intro}
Hot Jupiters have been subjected to intensive studies of 
their orbits
and atmospheres. When one imagines a ``typical" hot Jupiter, one would think of
a jovian planet orbiting a relatively metal-rich solar-type star within 3 days. 
Many other characteristics of hot Jupiters have been discussed in the literature, 
including the inflation of their radii
when the insolation from the central stars becomes strong \citep[e.g.,][]{2010exop.book..397F}.
Hot Jupiters, at least around relatively cool stars ($T_{\mathrm{eff}} \lesssim 6200$ K), generally have circular orbits aligned with their host stars'
spin axes \citep[][]{2010ApJ...718L.145W, 2012ApJ...757...18A}. 
Moreover, hot Jupiters are generally isolated up to a certain distance from their host stars \citep{2012PNAS..109.7982S} with the exception of WASP-47 
\citep{2015ApJ...812L..18B}, but are likely to have some ``friend(s)" at longer orbital separations. 
These friends are outer planetary and/or stellar companions, and have been revealed by long-term RV monitoring and high-contrast imaging campaigns \citep[e.g.,][]{2014ApJ...785..126K, 2015ApJ...800..138N, 2015ESS.....310106N}.
However, such intensive studies have not settled the serious issues for both the origin of hot Jupiters and their properties, 
and we have not reached a consensus for the formation and evolution of hot Jupiters. \par

The {\it Kepler} satellite's second mission, K2, 
has provided new opportunities to 
search and characterize transiting planets including hot Jupiters. 
K2 has already discovered many outstanding systems including 
possibly rocky planets around cool dwarfs \citep{2015ApJ...804...10C, 2015ApJ...811..102P}
and a disintegrating minor planet around a
white dwarf \citep{2015Natur.526..546V}. 
To find and characterize the unique planetary systems
detected by K2, we initiated the new collaboration ESPRINT, {\it 
Equipo de Seguimiento de Planetas Rocosos Intepretando sus Transitos}, 
which has already confirmed/validated 
a disintegrating rocky planet \citep[ESPRINT I:][]{2015ApJ...812..112S}, 
small planets around solar-type stars 
\citep[ESPRINT II:][]{2016ApJ...820...56V} and
a super-Earth/mini-Neptune around a mid-M dwarf, 
for which intensive follow-up studies are expected 
\citep[ESPRINT III:][]{2016ApJ...820...41H}. 

In this paper, we report the discovery and early characterization of
a hot Jupiter, detected by our pipeline applied to the K2 Cycle 5 field stars. 
Our target, \target, is an F-type star with an effective temperature 
of $6200$ K,
as inferred from its colors. The top part of Table \ref{hyo1} summarizes the properties of \target\
as collected from the SDSS and 2MASS catalogs \citep{2012ApJS..203...21A, 
2006AJ....131.1163S}. Our earlier analysis implied that the planetary candidate,
\target b, orbits its central star with a period extremely close to 3 days. 
With intensive follow-up observations including radial velocity (RV) measurements 
and high-contrast imaging, we confirm that 
\target b is indeed a hot Jupiter orbiting \target.  
Our analysis suggests 
the hot Jupiter to be a typical, but still important, example of its class, exhibiting many of the properties described above.

The rest of the paper is organized as follows. We describe 
our pipeline to reduce the K2 data and detect planetary candidates in K2 field 5 (Section \ref{s:K2}); Section \ref{s:RVfollowup} describes our
follow-up campaign with RV measurements by the High Accuracy Radial velocity Planet Searcher North (HARPS-N) on the 3.6-m Telescopio Nazionale Galileo (TNG)
and High Dispersion Spectrograph (HDS) on the Subaru 8.2-m telescope,  
in which a complete spectroscopic transit of \target b is covered.  
Section \ref{s:hicontrast}
presents lucky imaging at 1.52-m Telescopio Carlos S\'anchez (TCS)
and adaptive-optics (AO) imaging with Subaru/HiCIAO.
We perform a simultaneous fit to the reduced K2 light curve and observed RVs by the two spectrographs, including the modeling of the Rossiter-McLaughlin (RM) effect \citep[e.g.,][]{2005ApJ...622.1118O, 2005ApJ...631.1215W} of \target b in Section \ref{s:analysis}. 
Section \ref{s:discussion} is devoted to discussion and summary. 

\section{Observations and Data Reductions}\label{s:obs}
\subsection{K2 Photometry}\label{s:K2}
\begin{figure*}[t]
\begin{center}
\includegraphics[width=16cm]{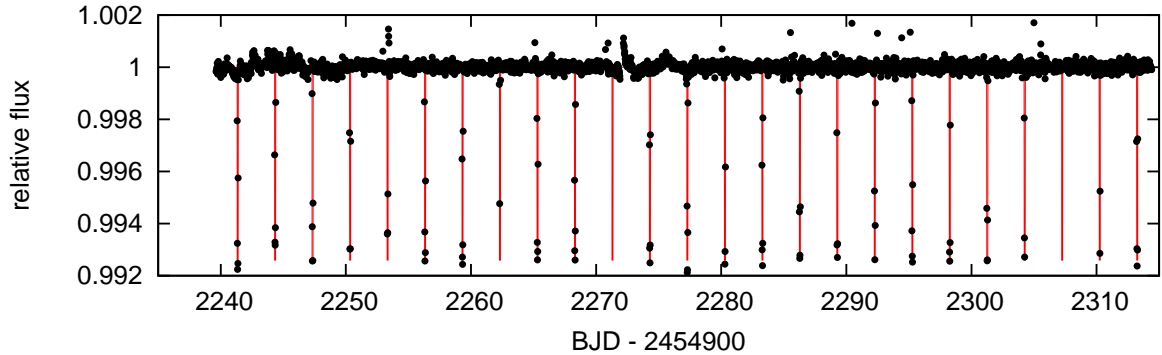} 
\caption{The full, reduced K2 light curve for \target. 
The equally-spaced red vertical lines correspond to the transits of \target b. }
\label{fig:fullcurve}
\end{center}
\end{figure*}

\target\ was observed from the $27^{\rm th}$ of April to the $10^{\rm th}$ of July 2015 as a pre-selected target star of {\it K2} Campaign 5. 
We downloaded the images of \target\ from the MAST website. The production of detrended  {\it K2} light curves by our collaboration was described 
in detail in the ESPRINT I paper \citep{2015ApJ...812..112S}. 
We then searched the light curves for transiting planet candidates with a Box-Least-Squares routine \citep{2002A&A...391..369K, 2010ApJ...713L..87J} using the optimal frequency sampling described by \citet{2014A&A...561A.138O}. \target\ was clearly detected with a signal-to-noise ratio (SNR) of 15.5. A linear ephemeris analysis of the individual transits yielded a best-fit period of $2.995637 \pm 0.000011$  days and a mid-transit time of $T_{c,0}=2457141.35116 \pm 0.00015$ (BJD).
Figure \ref{fig:fullcurve} plots the full reduced light curve; the deep transits (marked by red lines) are clearly visible.

\begin{table}[tb]
\begin{center}
\caption{Stellar Parameters of \target }\label{hyo1}
\begin{tabular}{lc}
\hline
Parameter & Value \\\hline\hline
\multicolumn{2}{l}{\it (Stellar Parameters from the SDSS and 2MASS Catalogs)} \\
RA & $08:30:18.91$ \\
Dec & $22:14:09.27$\\
$m_g$ (mag) & $11.876\pm 0.001$ \\
$m_r$ (mag) & $11.518\pm 0.001$ \\
$m_i$ (mag) & $11.407\pm 0.001$ \\
$m_z$ (mag) & $13.449\pm 0.014$ \\
$m_J$ (mag) & $10.528\pm0.025$ \\
$m_H$ (mag) & $10.258\pm0.022$ \\
$m_{K_\mathrm{s}}$ (mag) & $10.193\pm0.017$ \\
\hline
\multicolumn{2}{l}{\it (Spectroscopic Parameters)} \\
$T_{\rm eff}$ (K) & $6087\pm 38$ \\
$\log g$ (dex) & $4.106\pm0.075$\\
$[\mathrm{Fe/H}]$ (dex) & $0.15\pm0.04$ \\
$\xi$ (km s$^{-1}$) & $0.86\pm 0.19$\\
$V\sin I_s$ (km s$^{-1}$) & $5.65 \pm 0.33 $\\
$\zeta_\mathrm{RT}$ (km s$^{-1}$; assumed) & $4.5 \pm 0.7 $\\
\hline
\multicolumn{2}{l}{\it (Derived Parameters by Empirical Relation)} \\
$M_\star$ ($M_\odot$) & $1.31_{-0.09}^{+0.10}$ \\
$R_\star$ ($R_\odot$) & $1.66_{-0.17}^{+0.19}$\\
$\rho_\star$ ($\rho_\odot$) & $0.29_{-0.08}^{+0.10}$ \\
\hline
\multicolumn{2}{l}{\it (Derived Parameters by Y$^2$ Isochrone)} \\
$M_\star$ ($M_\odot$) & $1.37\pm 0.07$ \\
$R_\star$ ($R_\odot$) & $1.72_{-0.18}^{+0.20}$\\
$\rho_\star$ ($\rho_\odot$) & $0.27_{-0.07}^{+0.09}$ \\
age (Gyr) & $2.88_{-0.24}^{+0.26}$\\
\hline
\end{tabular}
\end{center}
\end{table}

\subsection{High Dispersion Spectroscopy}\label{s:RVfollowup}

\subsubsection{TNG/HARPS-N}

In order to confirm the planetary nature of \target b detected above, we observed \target\ with TNG/HARPS-N for precise RV measurements on 2015 November 18-25 UT as part of the observing program CAT15B\_79. HARPS-N \citep{2012SPIE.8446E..1VC} is a fiber-fed, echelle, thermally stable spectrograph in a vacuum, located on TNG at the Roque de los Muchachos Observatory (ORM) on La Palma, Spain. 
It covers the visible wavelength range between 383 nm and 693 nm with a resolving power of $R = 115,000$. TNG/HARPS-N RV measurements and their uncertainties were obtained with the G2 cross-correlation mask using the DRS pipeline, which is based on the weighted CCF method \citep{2002A&A...388..632P}. Thirteen RV measurements collected during the TNG/HARPS-N run allowed us to confirm the planetary nature of \target b, and to designate the target for a spectroscopic transit observation from Mauna Kea on November 27 in order to measure the RM effect of the system. Our TNG/HARPS-N measurements are presented in Table \ref{hyo2}.

\begin{table}[tb]
\begin{center}
\caption{RV Measurement by TNG/HARPS-N (absolute values)
}\label{hyo2}
\begin{tabular}{lrr}
\hline
BJD & value (m s$^{-1}$) & error (m s$^{-1}$) \\\hline\hline
$2457347.61743$ & $46554.8$ & $9.0$ \\
$2457347.64159$ & $46553.3$ & $8.5$ \\
$2457348.71636$ & $46194.6$ & $2.8$ \\
$2457348.73884$ & $46189.2$ & $2.2$ \\
$2457349.78322$ & $46489.3$ & $3.9$ \\
$2457349.79341$ & $46495.2$ & $4.6$ \\
$2457350.60492$ & $46540.6$ & $5.7$ \\
$2457350.69485$ & $46546.8$ & $3.4$ \\
$2457350.78737$ & $46491.6$ & $5.2$ \\
$2457351.66537$ & $46200.7$ & $6.3$ \\
$2457351.68768$ & $46177.3$ & $6.5$ \\
$2457352.60966$ & $46413.3$ & $9.8$ \\
$2457352.63088$ & $46446.9$ & $9.8$ \\
\hline
\end{tabular}
\end{center}
\end{table}

\subsubsection{Subaru/HDS}\label{s:hds}
We also observed \target\ with Subaru/HDS for precise RV measurements on 2015 
November 26-28 UT and 2016 February 2 UT.
We employed the standard I2a setup with Image Slicer \#2, 
covering the spectral region 
between $493-759$ nm with $R=85,000$ \citep{2012PASJ...64...77T}. 
For the precise RV measurements, we used the iodine (I$_2$) cell
and have the stellar light transmit through that cell to imprint the 
rich absorption lines of I$_2$ in the stellar spectra. 
On November 27, a spectroscopic transit was visible from Mauna Kea, and we took that
opportunity to measure the RM effect of the system, covering the whole transit
of \target b ($\sim$2 hours). 
On the same night, we obtained the stellar spectrum 
without the I$_2$ cell for a template in the RV analysis 
as well as for the detailed spectroscopic characterization of \target. 
During the twilight, we also obtained a flat-lamp spectrum transmitted through the I$_2$ cell
so that we can estimate the instrumental profile (IP) of HDS on that observing night. 
During the November run,
we had some clouds throughout the week on Mauna Kea, and the typical photon
counts of the raw spectra were about half of what we had initially expected. 

We reduced the raw data in a standard manner using the IRAF package
to extract the wavelength-calibrated, one-dimensional (1D) spectra. 
The signal-to-noise ratio (SNR) of the 1D spectra was typically 60 
per pixel. 
We then input the I$_2-$in spectra into the RV pipeline 
to extract the relative RV values with respect to the I$_2-$out 
template spectrum. 
The RV analysis for Subaru/HDS is described in detail in \citet{2002PASJ...54..873S,2012PASJ...64...97S}.  
The resulting RV values are listed in Table \ref{hyo3} along with their uncertainties. 

\begin{table}[tb]
\begin{center}
\caption{RV Measurement by Subaru/HDS. 
Only ``relative RVs" are obtained from the RV analysis with the I$_2$ cell, 
and values in this table are different from the absolute RVs listed
in Table \ref{hyo2}. 
}\label{hyo3}
\begin{tabular}{lrr}
\hline
BJD & value (m s$^{-1}$) & error (m s$^{-1}$) \\\hline\hline
$2457353.13038$ & $208.2$ & $9.8$ \\
$2457353.99119$ & $48.4$ & $9.7$ \\
$2457353.99899$ & $43.3$ & $10.1$ \\
$2457354.00678$ & $57.5$ & $11.4$ \\
$2457354.01458$ & $38.4$ & $10.6$ \\
$2457354.02240$ & $34.3$ & $10.4$ \\
$2457354.03020$ & $35.1$ & $11.3$ \\
$2457354.03800$ & $27.1$ & $10.0$ \\
$2457354.04580$ & $27.0$ & $10.7$ \\
$2457354.05360$ & $14.2$ & $9.8$ \\
$2457354.06139$ & $-6.0$ & $9.7$ \\
$2457354.06919$ & $-20.6$ & $10.5$ \\
$2457354.07699$ & $-12.0$ & $10.7$ \\
$2457354.08479$ & $-5.5$ & $11.7$ \\
$2457354.09258$ & $-11.4$ & $11.1$ \\
$2457354.10039$ & $14.1$ & $10.5$ \\
$2457354.97403$ & $-156.2$ & $10.3$ \\
$2457420.91151$ & $-186.8$ & $10.8$ \\
$2457420.91917$ & $-184.4$ & $11.2$ \\
$2457420.92683$ & $-163.2$ & $11.0$ \\
\hline
\end{tabular}
\end{center}
\end{table}
We measured the equivalent widths for \ion{Fe}{1} and \ion{Fe}{2} lines of the I$_2-$out template spectrum, making it possible to estimate the atmospheric parameters of \target. 
As described in \citet{2002PASJ...54..451T, 2005PASJ...57...27T}, 
we estimated the stellar effective temperature $T_\mathrm{eff}$, surface gravity $\log g$, metallicity [Fe/H],
and microturbulent velocity $\xi$ from the excitation and ionization equilibria. 
We also measured the projected rotational velocity of the star by convolving a theoretically
synthesized spectrum with the rotation plus macroturbulence broadening kernel 
\citep[the radial-tangential model:][]{2005oasp.book.....G} 
and the IP of HDS. 
Following \citet{2012ApJ...756...66H, 2014ApJ...783....9H}, 
we adopted the empirical relation by \citet{2005ApJS..159..141V} for the macroturbulent velocity $\zeta_\mathrm{RT}$ as
\begin{eqnarray}
\zeta_\mathrm{RT}
=\left(3.98+\frac{T_\mathrm{eff}-5770~\mathrm{K}}{650~\mathrm{K}}
\right)~\mathrm{km~s}^{-1}. 
\label{zetaRT}
\end{eqnarray}
The results of these measurements are summarized in the second part of Table \ref{hyo1}.

To estimate the physical parameters of the star, we converted the atmospheric parameters into mass and radius (and density), employing two different methods. 
We first used the empirical relations for stellar mass and radius derived by \citet{2010A&ARv..18...67T}, which is based on physical parameters measured using detached binaries. 
We also converted the atmospheric parameters using the Yonsei-Yale (Y$^2$)
isochrone model \citep{2001ApJS..136..417Y}. 
For both cases, we ran Monte Carlo simulations to estimate the uncertainties for the physical parameters by randomly generating the sets of atmospheric parameters assuming Gaussian errors. 
Since the uncertainties for atmospheric parameters listed in Table \ref{hyo1} are all statistical errors, we quadratically added a systematic error of $40$ K for the effective temperature to account for the systematics estimated by \citet{2010MNRAS.405.1907B}, who compared the luminosity-based effective temperatures with the spectroscopically measured values as presented here. 
The resulting parameters, including their uncertainties, are summarized in the bottom parts of Table \ref{hyo1}. 
The mass and radius estimated by the two techniques are consistent with 
each other within $1\sigma$. Since the uncertainties for the empirical mass
and radius take into account the systematic errors in the empirical 
relation, we use those estimates for the rest of this paper.

\subsection{High Contrast Imaging} \label{s:hicontrast}

We obtained high-contrast images of \target\ to search for stellar companions, and to exclude the possibility that the transit signal is a false positive from a background eclipsing binary.  Our data, described below, consist of lucky imaging observations with the FastCam camera and adaptive optics imaging with HiCIAO on the Subaru Telescope.

\subsubsection{TCS/FastCam Lucky Imaging Observations}

On 2015 November 15 UT, 20,000 individual frames of \target\ were collected in the $I-$band using FastCam \citep{2008SPIE.7014E..47O} at TCS in the Observatorio del Teide, Tenerife, with 50 ms exposure time for each frame. FastCam is an optical imager with a low noise EMCCD camera which allows to obtain speckle-featuring non-saturated images at a fast frame rate
\citep[see][]{2011A&A...526A.144L}.

\begin{figure}
\centerline{\includegraphics[angle=0,scale=0.425]{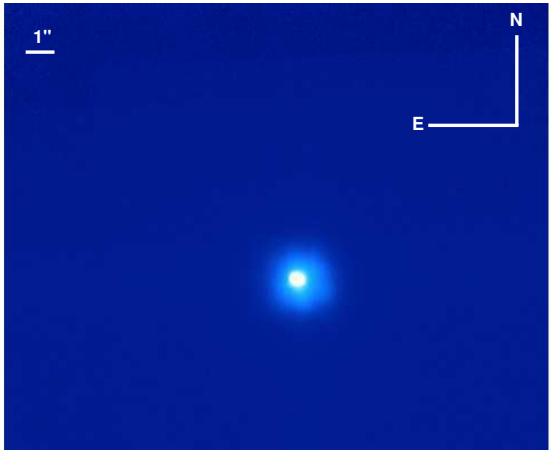}}
\caption{Final image of \target\ after lucky imaging processing with a 30\% selection of the best individual TCS/FastCam frames.\label{fig-EPIC212110888-tcs_fastcam-li-20151115-30percent_a}}
\end{figure}

In order to construct a high resolution, diffraction limited, long-exposure image, the individual frames were bias subtracted, aligned and co-added using a Lucky Imaging (LI) algorithm \citep[see][]{2006A&A...446..739L}. The LI selection is based on the brightest speckle in each frame, which has the highest concentration of energy and represents a diffraction limited image of the source. Those frames with a larger count number at the brightest speckle are the best ones. The percentage of the best frames chosen depends on the natural seeing conditions and the telescope diameter and is based on a trade between a sufficiently high integration time, given by a higher percentage, and a good angular resolution, obtained by co-adding a lower amount of frames. Figure \ref{fig-EPIC212110888-tcs_fastcam-li-20151115-30percent_a} presents the high resolution image constructed by co-addition of the 30\% of the best frames, i.e. with 300-second total exposure time. The combined image achieved $\Delta m_I=3.8-4.0$ at $1\farcs0$, and no bright companion was detectable in the images within $1\farcs0$.

\subsubsection{Subaru/HiCIAO Observations} \label{s:hiciao}

We used adaptive optics imaging on the Subaru Telescope to rule out the presence of a background eclipsing binary at smaller angular separations and to search for faint stellar companions.  Our observations were conducted on 2015 December 30 UT using the adaptive optics system AO188 \citep{Hayano_2010_AO188} and the high-contrast near-infrared camera HiCIAO \citep{Suzuki_HiCIAO_2010}.  Using \target\ as a natural-guide star, we acquired 60 dithered frames with individual integration times of 15 s for a total integration time of 900 s. 
The point spread functions (PSFs) of the primary star in these images were intentionally saturated within typically $0\farcs07$, to search for faint companions.
We also took unsaturated images with a 9.74\% neutral-density filter to verify the star's position and flux.  We obtained images of the globular cluster M5 for astrometric calibration.

\subsubsection{HiCIAO Data Reductions} \label{s:hiciao_reduc}

We reduced our HiCIAO data using the ACORNS pipeline, described in \cite{Brandt+13_ACORNS}.  We removed correlated read noise, masked hot pixels, flat-fielded, and corrected instrumental distortion by comparing images of the globular cluster M5 with archival data from the {\it Hubble Space Telescope}.  We then aligned the images, but did not apply high-contrast algorithms to suppress diffracted starlight. \par

\begin{figure}
\centering\includegraphics[width=\columnwidth]{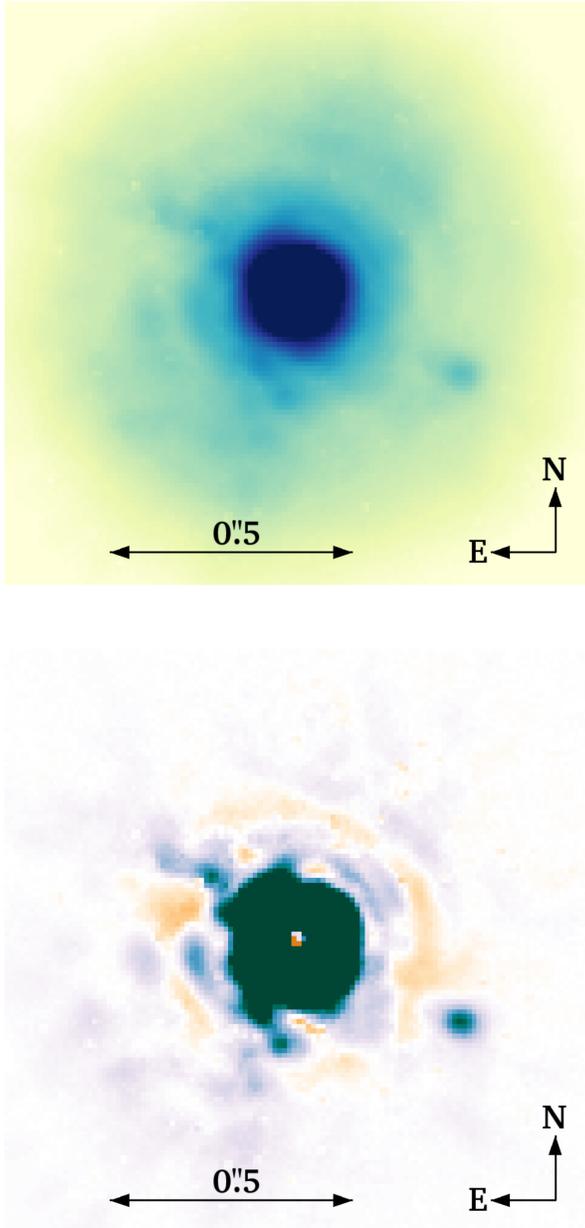}
\caption{HiCIAO $H$-band image of~\target~showing its companion candidate (CC). {\it Top}:  Combined image before applying a high-pass filter; the stretch is logarithmic. A faint CC can be seen at a projected separation of 0\farcs36 to the southwest.  {\it Bottom}:  Combined image after the high-pass filter; the stretch is linear. The CC is nearly 300 times fainter in $H$ band than the primary star. }
\label{fig_image}
\end{figure}

We combined all the saturated images 
to find a faint companion candidate (CC) around \target\ (see Figure \ref{fig_image}).  However, the CC appears to be embedded in the bright halo of primary star's PSF.  To suppress the halo, we applied high-pass filter to each saturated image after the image registration.  We used a median filter with a width of 4 PSF full widths at half maximum (FWHM; 1 FWHM $\approx$ 54 mas), subtracting the filtered image from each of our original frames.  The lower panel of Figure \ref{fig_image} displays the final combined image, on which the CC is clearly detected.  We measure a centroid of CC and estimate the separation and position angles between the CC and primary star to be 361.3 $\pm$ 3.5 mas and 206\fdg77 $\pm$ 0\fdg62, respectively. The CC's position was measured in the frames where the primary star's PSF is saturated. We corrected the primary star's centroids using the unsaturated frames whose acquisitions were interspersed through the data-acquisition sequence for the saturated frames.\par

We performed aperture photometry for the CC on the combined, high-pass-filtered image shown in the bottom panel of Figure \ref{fig_image}. 
It is notable that the high-pass filter decreases the flux of CC.  We estimated and recovered the flux loss ($\sim$35\%) based on the reductions for the images with injected artificial sources.  We measure a final, corrected $H$-band contrast of $\rm{\Delta}m_H = 6.19 \pm 0.11$~mag. 
This brightness contrast was derived using the primary star's flux in the unsaturated frames, for which a simultaneous photometry of the primary star and CC is prohibited. Then, the variation of AO correction is attributed to the photometry uncertainty, and we consider that the error term related to this variation is represented by the scatter of the photometry of unsaturated PSFs. 

\section{Global Analysis}\label{s:analysis}

\begin{figure}
\centering\includegraphics[width=\columnwidth]{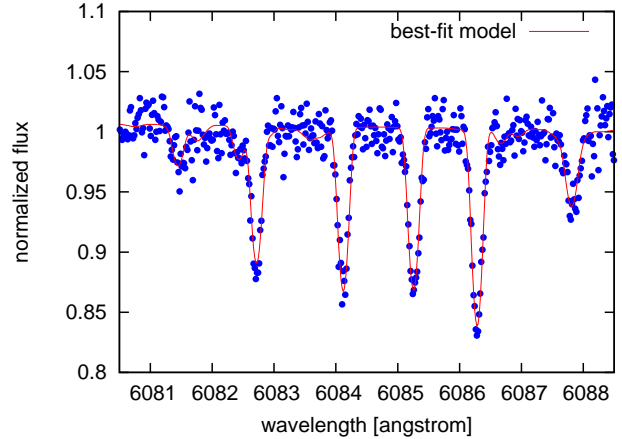}
\caption{Part of \target's spectrum taken by Subaru/HDS. 
No secondary peak is visible in the spectrum. }
\label{hds_spectrum}
\end{figure}

The high-contrast images by TCS/FastCam (in $I-$band) and Subaru/HiCIAO 
(in $H-$band) show no nearby source bright enough
to cause a transit-like signal as deep as $0.7\%$.  A background eclipsing binary, even if it achieved the maximum possible occulation (50\%), would need to be no more than $\sim$70 times ($\sim$4.6 magnitudes) fainter than \target.  Our observations, combined with SDSS archival images, confirm that no such sources exist between $0\farcs2$ and 20$^{\prime\prime}$.
It is still possible that a relatively bright binary companion
is present within $0\farcs2$ from \target, but visual inspection of 
the HDS spectrum did not show any secondary peak as shown in Figure \ref{hds_spectrum}.
Along with the fact that the observed RVs clearly show the 
presence of a planet-sized companion, whose RV variation is synchronous 
with the predicted phase from the K2 transits, 
we conclude that the periodic dimming seen in the K2 light curve 
is associated with a jovian planet orbiting \target.

Here, we attempt to simultaneously fit the K2 light curve with the RVs observed by HARPS-N and HDS.
The fitting procedure is similar in many aspects to those described in \citet{2015ApJ...799....9H} and \citet{2015ApJ...802...57S}. 
We use the following $\chi^2$ statistics in the global fit:
\begin{eqnarray}
\label{eq:chisq}
\chi^2 &=& \sum_{i}\frac{(f_\mathrm{LC,obs}^{(i)}-f_\mathrm{LC, model}^{(i)})^2}{\sigma_\mathrm{LC}^{(i)2}}\nonumber\\
&& + \sum_{i}\frac{(v_\mathrm{HARPS,obs}^{(i)}-v_\mathrm{HARPS, model}^{(i)})^2}{\sigma_\mathrm{HARPS}^{(i)2}}\nonumber\\
&& + \sum_{i}\frac{(v_\mathrm{HDS,obs}^{(i)}-v_\mathrm{HDS, model}^{(i)})^2}{\sigma_\mathrm{HDS}^{(i)2}},
\end{eqnarray}
where $f_\mathrm{LC,obs}^{(i)}$, $v_\mathrm{HDS,obs}^{(i)}$, and $v_\mathrm{HARPS,obs}^{(i)}$
are $i-$th observed K2 flux, HDS RV value, and HARPS RV value, 
and $\sigma_\mathrm{LC}^{(i)}$, $\sigma_\mathrm{HDS}^{(i)}$, and $\sigma_\mathrm{HARPS}^{(i)}$ are
their errors, respectively. 
To compute the model flux $f_\mathrm{LC, model}^{(i)}$ observed by K2, we integrate the 
transit model by \citet{2009ApJ...690....1O} over the cadence of the K2 observation ($\sim$29.4 minutes).

For the RV model, we adopt the following equations:
\begin{eqnarray}
\label{RVmodel}
v_\mathrm{HARPS, model}&=&K[\cos(f+\omega)+e\cos\omega]+\gamma_\mathrm{HARPS},\\
v_\mathrm{HDS, model}&=&K[\cos(f+\omega)+e\cos\omega]+\D v_\mathrm{RM}+\gamma_\mathrm{HDS},~~~
\end{eqnarray}
where $K$, $f$, $e$, and $\omega$, $\gamma_\mathrm{HDS}$, $\gamma_\mathrm{HARPS}$ 
are the RV semi-amplitude, true anomaly, orbital eccentricity, 
argument of periastron, and RV offsets for the HDS and HARPS data sets, 
respectively. Since the HDS data set covered a complete transit of \target b, 
we introduce the velocity anomaly term $\D v_\mathrm{RM}$ due to 
the RM effect for that data set only. 
We adopt the analytic formula by \citet{2011ApJ...742...69H},
in which $\D v_\mathrm{RM}$ is computed in terms of the projected rotational velocity $V\sin I_s$. For other spectroscopic parameters that appear in Equation (16) of 
\citet{2011ApJ...742...69H}, we assume the Gaussian and Lorentzian widths 
of $\beta=2.7$ km s$^{-1}$ and $\gamma=1.0$ km s$^{-1}$, and the macroturbulent
velocity of $\zeta_\mathrm{RT}=4.5$ km s$^{-1}$ as in Section \ref{s:hds}. 
We here neglect the convective blue-shift \citep{2011ApJ...733...30S}, 
since its impact is small enough ($\lesssim 3$ m s$^{-1}$ at the most) 
compared with the internal errors of Subaru/HDS RV data ($\sim$10~m~s$^{-1}$).

Assuming that the likelihood is proportional to $\exp(-\chi^2/2)$, 
we run a Markov Chain Monte Carlo (MCMC) simulation to estimate 
the global posterior distribution of the fitting parameters. 
The fitting parameters in our model are orbital period $P$, 
time of the transit center $T_{c,0}$, scaled semi-major axis $a/R_s$, 
transit impact parameter $b$, planet-to-star radius ratio $R_p/R_s$, 
and limb-darkening parameters $u_1+u_2$ and $u_1-u_2$ for the K2 
data set assuming a quadratic law, $e\cos\omega$, $e\sin\omega$, 
$K$, $V\sin I_s$, the projected obliquity $\lambda$,
$\gamma_\mathrm{HDS}$, and $\gamma_\mathrm{HARPS}$. 
Among these, $P$, $T_{c,0}$, $a/R_s$, $b$, $R_p/R_s$, $e\cos\omega$, 
$e\sin\omega$, are related to both the light curve and RV
data sets, but the others are for RV data only 
(except the limb-darkening coefficients). 
Since limb-darkening coefficients 
are weakly constrained from the K2 transit curve, we put weak Gaussian 
priors on $u_1+u_2$ and $u_1-u_2$, 
with their centers being 0.65 and 0.08, respectively,
and dispersions of $0.20$, based on the theoretical 
values by \citet{2011A&A...529A..75C} for the Kepler band. 
The RM velocity anomaly $\D v_\mathrm{RM}$ also weakly depends 
on the limb-darkening coefficients. 
The RV precision and sparse time sampling of the Subaru/HDS dataset,
however, make the fit of those coefficients almost impossible, 
and we opted to fix them at $u_{1,\mathrm{RM}}=0.43$ and $u_{2,\mathrm{RM}}=0.28$ based on the theoretical values for the $V-$band \citep{2011A&A...529A..75C}. 
We note that we allow the orbital period to vary rather than fix
it at the value from the light curve analysis alone. In this way,
the ephemeris of \target b is globally determined from the light curve
and RV data, and the uncertainty in modeling the RM effect 
reflects the uncertainty of the period.

\begin{figure}[t]
\begin{center}
\includegraphics[width=8.5cm]{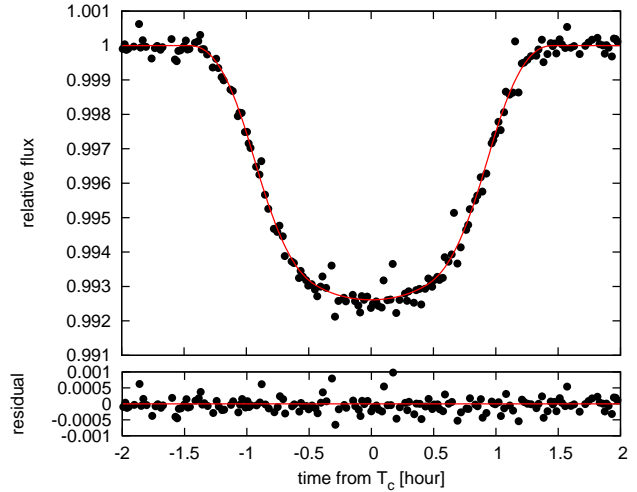} 
\caption{Phase-folded K2 light curve around the transit
of \target b (black points). The best-fit model is 
plotted by the red solid line. 
Bottom panel indicates the residual between the two. 
}
\label{fig:K2}
\end{center}
\end{figure}

\begin{figure}[t]
\begin{center}
\includegraphics[width=8.5cm]{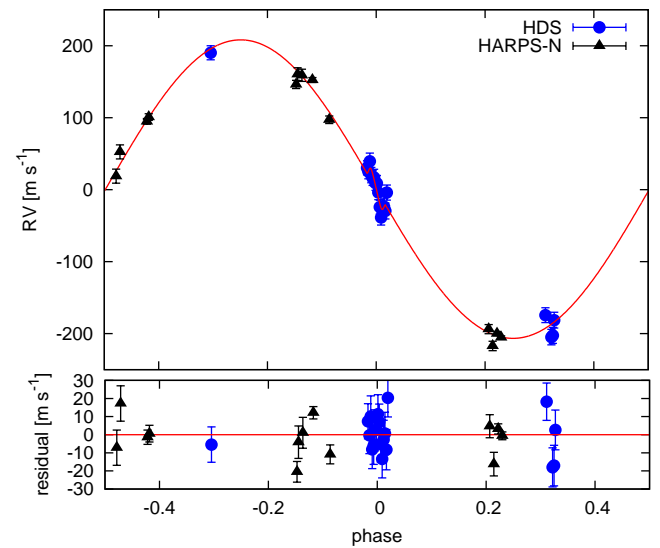} 
\caption{Phase-folded RV variation by Subaru/HDS (blue circles)
and TNG/HARPS-N (black triangles). 
The best-fit RV offset in each dataset ($\gamma_\mathrm{HARPS}$ and 
$\gamma_\mathrm{HDS}$) is subtracted in the plot. 
The best-fit model is plotted
by the red solid line. Bottom panel indicates the residual between
the two. 
}
\label{fig:orbit}
\end{center}
\end{figure}

\begin{figure}[t]
\begin{center}
\includegraphics[width=8.5cm]{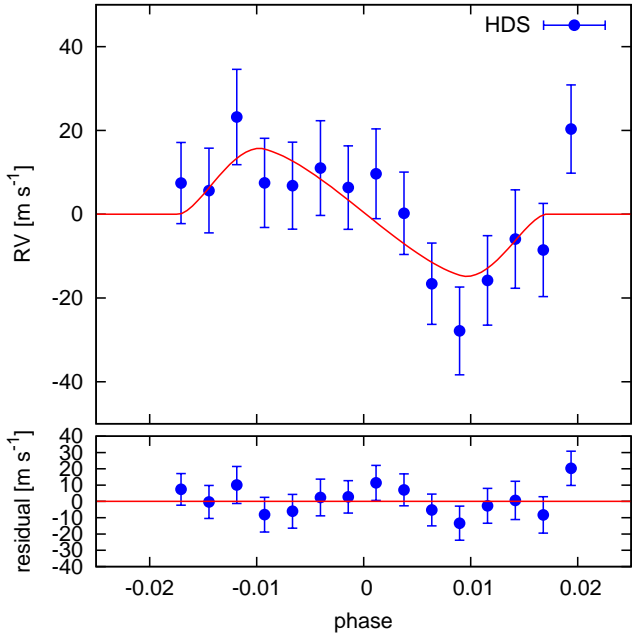} 
\caption{RV variation around the transit of \target b after
subtracting the orbital motion. The best-fit model is plotted
by the red solid line. Bottom panel indicates the residual between
the two. 
}
\label{fig:RM}
\end{center}
\end{figure}

We use our customized code \citep[e.g.,][]{2015ApJ...799....9H} to perform the global fit, in which Equation (\ref{eq:chisq}) is first minimized by the Nelder-Mead simplex method 
\citep[e.g.,][]{2002nrc..book.....P} and then the step size for each parameter is iteratively optimized, 
before running $10^7$ MCMC steps to estimate the global posterior. 
Since the accurate uncertainty for each flux value in K2 data is difficult to infer, we scaled 
$\sigma_\mathrm{LC}^{(i)}$ so that the reduced $\chi^2$ for the K2 data set becomes
approximately unity. 
We take the median, 15.87 and 84.13 percentiles of the maginalized posterior for each fitting
parameter to provide the best-fit value and its uncertainties, which are listed in Table \ref{hyo4}. 
The observed data along with the best-fit models are displayed 
in Figures \ref{fig:K2}, \ref{fig:orbit}, and \ref{fig:RM}, for
the phase-folded K2 light curve, orbital RVs, and RM velocity anomaly,
respectively. 

The best-fit model indicates that \target b is a typical
hot Jupiter in a 3-day, prograde orbit ($\lambda=-1_{-9}^{+10}$ degrees). 
Based on the best-fit model parameters in the global fit as
well as \target's physical parameters, we also compute 
\target b's physical and orbital parameters, including the planet
mass $M_p$, radius $R_p$, density $\rho_p$, orbital inclination
$i_o$, and semi-major axis $a$. 
The result is also summarized in Table \ref{hyo4}; the planet is consistent with a slightly inflated jovian planet, in a circular orbit (within $1\sigma$). The stellar density from the transit curve alone is
estimated to be $\rho_\star/\rho_\odot=0.448_{-0.056}^{+0.086}$,
which agrees with the spectroscopically measured stellar density
(Table \ref{hyo1}) with $1.4\sigma$, reinforcing the idea
that \target b is indeed transiting the F star.

The residual of the RM velocity anomaly in Figure \ref{fig:RM} 
seems to exhibit
a possible time-correlated noise, where each RV residual could be
correlated with the adjacent ones. 
To test if this is the case or not, we computed Pearson's 
correlation coefficient $r_0$ between the adjacent RV residual
values in Figure \ref{fig:RM}. 
We then ran a Monte Carlo simulation of $10^6$ steps
to estimate its $p-$value, where
we permutated the RV residuals randomly and recorded each 
correlation coefficient $r$ between the adjacent RV residuals 
for each dataset (step). 
Consequently, we obtained $r_0=-0.0427$ and found that its $p-$value
is high enough ($p(|r|>|r_0|)=0.876$), implying that the observed RV 
residuals have no significant correlation with the next ones
(i.e., time-correlated noise was not observed).

We note that $\lambda$ is a projected obliquity and the true, 
3-dimensional (3D) spin-orbit angle is not known; 
$\lambda\approx 0\arcdeg$ does not necessarily mean that the 
orbit is aligned. One way to break this degeneracy is to measure the stellar
inclination $I_s$, which is defined as the angle between the line-of-sight and 
stellar spin axis by the combination of the stellar rotation period, radius, and
projected rotation velocity $V\sin I_s$ \citep[e.g.,][]{2012ApJ...756...66H}.  
We inspected the periodogram of \target's light curve and found that 
there was a peak around $P=19$ days, which could be ascribed to 
the rotation period of \target, which is also expected 
by scaling the Sun's rotation period ($24.5$ days)
by $\sqrt{2.9\mathrm{Gyr}/4.6\mathrm{Gyr}}$. 
If this peak indeed corresponds to the period of rotation, 
the rotation velocity at the stellar equation is estimated as $\approx 4.4\pm 0.5$ km s$^{-1}$ based on the $R_s$ listed in Table \ref{hyo1}. Comparing this value with the observed $V\sin I_s$, we expect that $I_s$ is close to $90^\circ$, suggesting a small 3D spin-orbit angle.

\begin{table}[tb]
\begin{center}
\caption{Result of the Global Fit}\label{hyo4}
\begin{tabular}{lc}
\hline
Parameter & Value \\\hline\hline
{\it (Fitting Parameters)} & \\
$P$ (days) & $2.995654\pm 0.000018$ \\
$T_{c,0}$ (BJD) & $2457141.35087 \pm 0.00025$\\
$a/R_s$ & $6.70_{-0.29}^{+0.40}$ \\
$b$ & $0.822_{-0.034}^{+0.022}$ \\
$R_p/R_s$ & $0.0887_{-0.0012}^{+0.0009}$ \\
$u_1+u_2$ & $0.66\pm 0.12$ \\
$u_1-u_2$ & $0.07\pm 0.20$ \\
$e\cos\omega$ & $0.0037_{-0.0063}^{+0.0060}$ \\
$e\sin\omega$ & $-0.001\pm 0.014$\\
$K$ (m s$^{-1}$) & $207.3_{-2.2}^{+2.3}$\\
$V\sin I_s$ (km s$^{-1}$) & $5.0_{-1.4}^{+1.3}$\\
$\lambda$ ($^\circ$)& $-1_{-9}^{+10}$\\
$\gamma_\mathrm{HDS}$  (m s$^{-1}$) & $18.1\pm3.4$ \\
$\gamma_\mathrm{HARPS}$  (m s$^{-1}$) & $46394.4\pm 1.4$ \\
\hline
{\it (Derived Parameters)} & \\
$M_p$ ($M_J$) & $1.773\pm 0.086$ \\
$R_p$ ($R_J$) & $1.44\pm 0.16$ \\
$\rho_p$ ($\rho_J$) & $0.60_{-0.16}^{+0.25}$ \\
$i_o$ ($^\circ$) & $82.96_{-0.55}^{+0.69}$ \\
$a$ (AU) & $0.0445_{-0.0011}^{+0.0010}$ \\
$e$ & $<0.022$ \\
\hline
\end{tabular}
\end{center}
\end{table}

\section{Discussion and Summary }\label{s:discussion}
We have conducted intensive follow-up observations for a hot Jupiter candidate,
\target b, which was detected by our pipeline in an analysis of K2 field 5 stars. 
Our RV follow-up, along with the absence of a bright nearby source ($\Delta m<6$) in the high-contrast images, 
confirm the planetary nature of \target; we have determined the mass and radius 
of the planet to be $M_p=1.773\pm 0.086M_J$ and $R_p=1.44\pm 0.16R_J$,
respectively. Its central star, \target, is a relatively metal-rich, 
F-type star, which is typical of hot-Jupiter hosts.

We detected the velocity anomaly during the transit observed by HDS on 
November 27, with $3.6\sigma$ significance. Modeling of the RM effect implies 
that the orbit of \target b is prograde with respect to the stellar spin;
we estimate the best-fit value for the projected obliquity as $\lambda=-1_{-9}^{+10}$ degrees. 
To verify our result, 
we also tested the global fit with a Gaussian prior distribution 
for $V\sin I_s$ based on the spectroscopically measured value, 
and obtained a fully consistent result 
($\lambda=-1.1\pm 8.1$ degrees). 
This small obliquity is consistent with the well-known
finding that stars cooler than $T_\mathrm{eff}= 6250$ K 
generally have a small obliquity, while hotter ones tend to be misaligned
\citep[e.g.,][]{2010ApJ...718L.145W}. 
\target's effective temperature is $T_\mathrm{eff}=6087\pm 38$ K, which
is close to the alignment/misalignment divide, making it an important 
sample for future statistical analyses on observed obliquities.

\begin{figure}[t]
\begin{center}
\includegraphics[width=8.5cm]{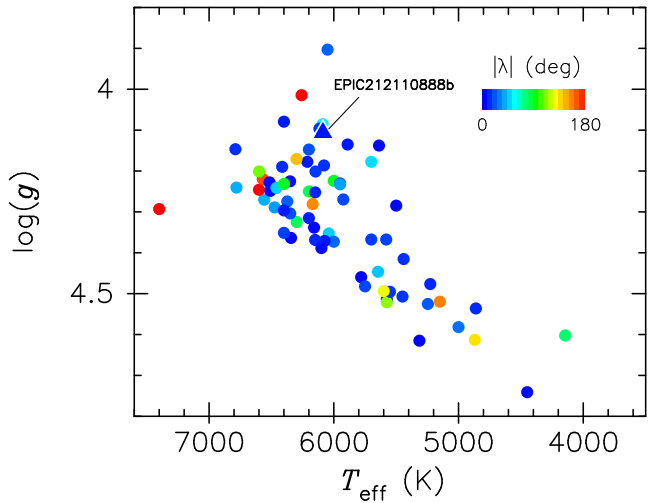} 
\caption{
Observed obliquities as a function of stellar temperature 
$T_\mathrm{eff}$ and surface gravity $\log g$. 
The values of $\lambda$, $T_\mathrm{eff}$, and $\log g$ 
were taken from 
http://www.astro.physik.uni-goettingen.de/\~{}rheller/
and http://exoplanets.org/. 
}
\label{fig:HR-lambda}
\end{center}
\end{figure}
\target\ is located around an edge of the main sequence in
the $T_\mathrm{eff}-\log g$ plane (Figure \ref{fig:HR-lambda}). 
This region of stellar evolution has fewer measurements
of the stellar obliquity. 
One possible channel for the formation of hot Jupiters is dynamical
processes (e.g., planet-planet scatterings) followed by tidal 
interactions between planets and their hosts 
\citep[e.g.,][]{2011ApJ...742...72N,2007ApJ...669.1298F}. 
Tidal interactions also tend to damp the stellar obliquity, 
but the precise timescale of this obliquity damping is not known 
and believed to depend on the stellar type \citep{2010ApJ...718L.145W,2014ApJ...784...66X}. 
In this context,
a comparison between the timescales for the obliquity damping by tides
and actual system ages could become an important clue. 
Considering that ages are generally better constrained for the slightly
evolved, but still hot stars such as \target\ than for their cooler counterparts, 
more obliquity measurements for those stars will 
provide additional insight into the tidal evolution of hot Jupiters.

It would be also of interest to discuss the observed obliquity
in terms of host star's mass. \citet{2015ApJ...811...82S} described 
that an alternative explanation for the observed trend of
stellar obliquity is that magnetic star-disk torques act to damp non-zero
stellar obliquities of less massive stars. Since the magnetic 
field of massive stars ($M_s\gtrsim 1.2M_\odot$) is much weaker by an 
order of magnitude, the stellar obliquity, which could be primordially
enhanced by e.g., an outer companion, is preserved and
a spin-orbit misalignment is likely observed around those stars
\citep[see Figure 1 in][]{2015ApJ...811...82S}. Provided that 
\target's mass is $1.31-1.37M_\odot$, the low obliquity in this system 
could be a new exception to the observed dependence of stellar obliquity
on the stellar mass.

AO imaging using Subaru/HiCIAO has revealed a possible companion with $\Delta m_H=6.19$ at a separation of $0\farcs36$. If this faint source is indeed bound to \target, the projected distance between the two components is estimated as $\sim 200$ AU assuming a distance to \target\ of $\sim 590$ pc based on its apparent and estimated absolute magnitudes. One can also estimate the absolute magnitude of the hypothetical stellar companion as $M_H=8.4$ mag using an isochrone \citep[e.g.,][]{2008ApJS..178...89D}, which corresponds to a mid-M dwarf whose mass is $\sim 0.2 M_\odot$.
The absolute Kepler magnitude of this CC is also 
estimated as $M_{Kp}=11-12$ mag. Considering \target's absolute magnitude 
of $M_{Kp}=3.0$, we can safely neglect the impact of dilution in the 
K2 transit curve ($\lesssim5\times 10^{-4}$). 
If this CC is indeed bound to \target, 
it also satisfies a general trend that hot Jupiters have outer giant planets and/or stellar companions \citep[e.g.,][]{2014ApJ...785..126K,2015ApJ...800..138N,2015ESS.....310106N}, but future follow-up observations are required to prove the physical association by checking the common proper motion and/or detecting the CC in a different observing band.

The relative brightness of the host star (e.g., $m_r=11.52$ mag)
makes \target b a good target for further follow-up to characterize its atmosphere. With the period so close to 3 days ($2.995654\pm 0.000018$
days), however, transit follow-ups from the ground are only possible 
around a certain longitude on the Earth for long intervals. This in turn means
that its transits are visible every 3 days at certain observatories, and
unusually accurate characterization may be possible through repeated observations of transits.
\vspace{3mm}\\
{\it Note: After completing the work described herein, we became aware of the independent discovery and 
characterization of \target b by \citet{2016arXiv160107635L}. 
Our measurements of the stellar and planetary properties are in agreement with theirs.}

\par
{\it Facilities:} \facility{Subaru (HDS, HiCIAO)}, \facility{TNG (HARPS-N)}, \facility{TCS (FastCam)}

\acknowledgments 

This paper is based on observations made with the Italian Telescopio Nazionale Galileo (TNG) operated on the island of La Palma by the Fundaci\'on Galileo Galilei of the INAF (Istituto Nazionale di Astrofisica) at the Spanish Observatorio del Roque de los Muchachos of the Instituto de Astrofísica de Canarias. We also provide observations with the 1.52-m Carlos Sanchez Telescope operated on the island of Tenerife by the Instituto de Astrof\'isica de Canarias in the Spanish Observatorio del Teide.
This paper is also based on data collected at Subaru Telescope, which is
operated by the National Astronomical Observatory of Japan.  
We acknowledge the support for our Subaru HDS observations by Akito
Tajitsu and Kentaro Aoki, support scientists for the Subaru/HDS. 
We also acknowledge the support by Hikaru Nagumo, Jun Hashimoto, and Jun-Ichi Morino for our Subaru observations, and David Lafreni\'{e}re, who generously provided us  his source code for ADI data reductions.
The data analysis was in part carried out on common use data analysis computer system at the Astronomy Data Center, ADC, of the National Astronomical Observatory of Japan.  PyFITS and PyRAF were useful for our data reductions.  PyFITS and PyRAF are products of the Space Telescope Science Institute, which is operated by AURA for NASA.  
Our analysis is also based on observations made with the NASA/ESA Hubble Space Telescope, and obtained from the Hubble Legacy Archive, which is a collaboration between the Space Telescope Science Institute, the Space Telescope European Coordinating Facility (ST-ECF/ESA) and the Canadian Astronomy Data Centre (CADC/NRC/CSA).
T.H.\ and M.K.\ are supported by Japan Society for Promotion of Science (JSPS) Fellowship for Research (No. 25-3183 and 25-8826). A.F.\ acknowledges support by the Astrobiology Center Project of National Institutes of Natural Sciences (NINS) (Grant Number AB271009).  This work was performed, in part, under contract with the Jet Propulsion Laboratory (JPL) funded by NASA through the Sagan Fellowship Program executed by the NASA Exoplanet Science Institute. 
N.N.\ acknowledges support by the NAOJ Fellowship, Inoue Science Research Award, and Grant-in-Aid for Scientific Research (A) (No. 25247026) 
from the Ministry of Education, Culture, Sports, Science and Technology (MEXT) of Japan. M.T.\ acknowledges support by
Grant-in-Aid for Scientific Research (No.15H02063). 
G.N., E.P., S.V., J.P.A., and I.R.\ acknowledge support from the Spanish Ministry of Economy and Competitiveness (MINECO) 
and the Fondo Europeo de Desarrollo Regional (FEDER) through grants ESP2013-48391-C4-1-R and ESP2014-57495-C2-2-R.
We acknowledge the very significant cultural role and reverence that the
summit of Mauna Kea has always had within the indigenous people in Hawai'i. 




\end{document}